\documentclass[bibyear]{aa} 
\usepackage{epsfig}
\usepackage{amsmath}
\usepackage{subfigure}
\usepackage{natbib}
\usepackage{multirow}
\usepackage{color}
\usepackage{longtable}
\usepackage{graphicx}
\usepackage{epstopdf}
\usepackage{booktabs}
\usepackage{txfonts}

\newcommand{\jmst}{J.~Mol.~Struct.}   

\newcommand{\jms}{J.~Mol.~Spectrosc.}

\newcommand{\kms}{km\,s$^{-1}$}
\begin{document}

\title{Cumulene carbenes in TMC-1: Astronomical discovery of $l$-H$_2$C$_5$ \thanks{Based on observations carried out
with the Yebes 40m telescope (projects 19A003, 20A014, and 20D15). The 40m radiotelescope at Yebes Observatory is operated by the Spanish Geographic Institute
(IGN, Ministerio de Transportes, Movilidad y Agenda Urbana).}}

\author{
C.~Cabezas\inst{1},
B.~Tercero\inst{2,3},
M.~Ag\'undez\inst{1},
N.~Marcelino\inst{1},
J.~R.~Pardo\inst{1},
P.~de~Vicente\inst{3}
and
J.~Cernicharo\inst{1}
}

\institute{Grupo de Astrof\'isica Molecular, Instituto de F\'isica Fundamental (IFF-CSIC), C/ Serrano 121, 28006 Madrid, Spain.
\email carlos.cabezas@csic.es; jose.cernicharo@csic.es
\and Observatorio Astron\'omico Nacional (IGN), C/ Alfonso XII, 3, 28014, Madrid, Spain.
\and Centro de Desarrollos Tecnol\'ogicos, Observatorio de Yebes (IGN), 19141 Yebes, Guadalajara, Spain.
}

\date{Received; accepted}

\abstract{We report the first detection in space of the cumulene carbon chain $l$-H$_2$C$_5$. A total of eleven rotational transitions, with $J_{up}$ = 7-10 and $K_a$ = 0 and 1, were detected in TMC-1 in the 31.0-50.4 GHz range using the Yebes 40m radio telescope. We derive a column density of (1.8$\pm$0.5)$\times$10$^{10}$ cm$^{-2}$. In addition, we report observations of other cumulene carbenes detected previously in TMC-1, to compare their abundances with the newly detected cumulene carbene chain. We find that $l$-H$_2$C$_5$ is $\sim$4.0 times less abundant than the larger cumulene carbene $l$-H$_2$C$_6$, while it is $\sim$300 and $\sim$500 times less abundant than the shorter chains $l$-H$_2$C$_3$ and $l$-H$_2$C$_4$. We discuss the most likely gas-phase chemical routes to these cumulenes in TMC-1 and stress that chemical kinetics studies able to distinguish between different isomers are needed to shed light on the chemistry of C$_n$H$_2$ isomers with $n$\,$>$\,3.}

\keywords{ Astrochemistry
---  ISM: molecules
---  ISM: individual (TMC-1)
---  line: identification
---  molecular data}

\titlerunning{Discovery of H$_2$C$_5$ in TMC-1}
\authorrunning{Cabezas et al.}

\maketitle

\section{Introduction}

Cumulene carbenes are highly polar carbon-chains with elemental formula H$_2$C$_n$.
They contain consecutive carbon-carbon double bonds and two nonbonded electrons localized on the
terminal C atom. These species play major roles as reaction intermediates in combustion
and plasma processes and they are also of astrophysical interest since some of them have
been detected in interstellar and circumstellar environments. The first astronomical
discovery of the simplest cumulene carbene propadienylidene, $l$-H$_2$C$_3$, was carried out
by \citet{Cernicharo1991a} toward the cold dark cloud TMC-1. Butatrienylidene, $l$-H$_2$C$_4$,
was detected in the carbon-rich circumstellar envelope of IRC+10216 by \citet{Cernicharo1991b}
and in TMC-1 by \citet{Kawaguchi1991} and later, the larger cumulene carbene hexapentaenylidene,
$l$-H$_2$C$_6$, was detected in TMC-1, IRC+10216 and L1527 \citep{Langer1997,Guelin2000,Araki2017}.

Cumulene carbenes are metastable isomers, lying 0.5-0.6 eV above the most stable isomer, which is a polyacetylene (HC$_n$H) for an even $n$ or a cyclic structure when the number of carbon atoms is odd. Thus, their astronomical detection demonstrates how far from thermochemical equilibrium is the composition of interstellar clouds. The case of HNC, which is less stable than HCN by about 0.6 eV but as abundant as HCN \citep{Herbst1978}, illustrates this point as well.

Pentatetraenylidene, $l$-H$_2$C$_5$, is the cumulene carbene from the isomeric family
with formula C$_5$H$_2$. $l$-H$_2$C$_5$ is a high energy isomer and lies 0.564 eV above
the most stable isomers, the nonpolar pentadiynylidene (HCCCCH) and ethynyl
cyclopropenylidene ($c$-C$_3$HCCH), whose energy separation is very small $\thicksim$0.043
eV. Other two species are part of this isomeric family, HCCCHCC and $c$-C$_3$H$_2$C$_2$,
placed at 0.737 and 0.910 eV above the most stable forms, respectively \citep{Seburg1997}.
All these isomers, except the nonpolar HCCCCH, have been studied in the laboratory
\citep{Travers1997,McCarthy1997,Gottlieb1998} and, thus, their transition frequencies are well known. However, until very recently none of them has been observed in space. \citet{Cernicharo2021a} have reported the first identification in TMC-1 of the most stable isomer of this family, $c$-C$_3$HCCH, using a high sensitivity line survey
gathered with the Yebes 40m radio telescope. This achievement opens the door to the identification
of high energy isomers of this family in TMC-1.

In this Letter we report the first identification of the $l$-H$_2$C$_5$ cumulene carbene in space towards TMC-1 and a comparative study of the previously detected cumulene carbenes in this source. The derived column densities for these singular energetic species are interpreted by chemical
models and used to understand the chemical processes leading to their formation.

\section{Observations}

The data presented in this work are part of a deep spectral line survey in the Q band towards
TMC-1 ($\alpha_{J2000}=4^{\rm h} 41^{\rm  m} 41.9^{\rm s}$ and $\delta_{J2000}=+25^\circ 41'
27.0''$) that was performed at the Yebes 40m radio telescope between November 2019 and April 2021. The survey was done using new receivers, built
within the Nanocosmos project\footnote{\texttt{https://nanocosmos.iff.csic.es/}} consisting
of two HEMT cold amplifiers covering the 31.0-50.4 GHz band with horizontal and vertical
polarizations. Fast Fourier transform spectrometers (FFTSs) with $8\times2.5$ GHz bands per lineal polarization allow a simultaneous scan of a
bandwidth of 18 GHz at a spectral resolution of 38.15 kHz. This setup has been described before by \citet{Tercero2021}.

The observations were performed using the frequency switching technique with a frequency
throw of 10\,MHz or 8\,MHz (see, e.g., \citealt{Cernicharo2021b,Cernicharo2021c}). The
intensity scale, antenna temperature ($T_A^*$), for the two telescopes used in this work
was calibrated using two absorbers at different temperatures and the atmospheric
transmission model ATM \citep{Cernicharo1985, Pardo2001}.Different frequency coverages were observed, 31.08-49.52 GHz and 31.98-50.42 GHz, which permitted us to check that no spurious ghosts are produced in the down-conversion chain. The signal coming from the receiver is downconverted to 1-19.5 GHz, and then split into 8 bands with a coverage of 2.5 GHz each of which are analyzed by the FFTs.

Pointing and focus corrections were obtained
by observing strong SiO masers towards nearby
evolved stars (IKTau and UOri). Pointing errors were always within $2''-3''$.
Total calibration uncertainties have been adopted to be 10~\% based
on the observed repeatability of the line intensities between dif-
ferent observing runs. All data have been analyzed using
the GILDAS package\footnote{\texttt{http://www.iram.fr/IRAMFR/GILDAS}}.

\section{Results and discussion}

\subsection{Detection of $l$-H$_2$C$_5$ in TMC-1}
\label{h2c5}

\begin{figure*}
\centering
\includegraphics[angle=0,width=0.80\textwidth]{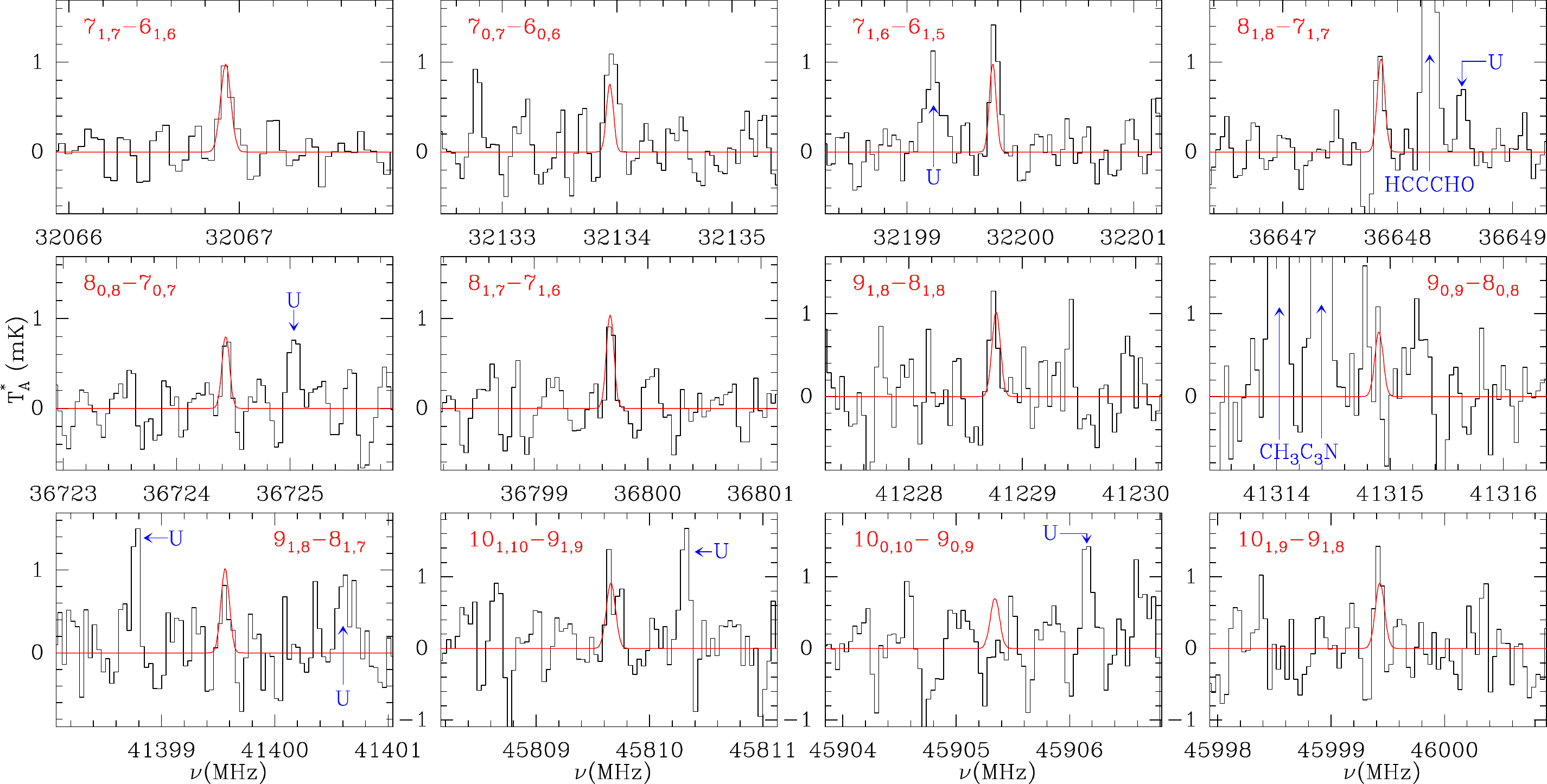}
\caption{Observed lines of $l$-H$_2$C$_5$ in TMC-1 in the 31.0-50.4 GHz range.
The abscissa corresponds to the rest frequency assuming a local standard of rest
velocity of 5.83\,\kms. The ordinate is antenna temperature in millikelvins. Curves shown in red are
the computed synthetic spectra. U labels correspond to features above 4$\sigma$. Frequencies and line parameters are given in Table \ref{table_fits}.} \label{spectra_h2c5}
\end{figure*}

$l$-H$_2$C$_5$ was observed in the laboratory by \citet{McCarthy1997} and the
prediction of its rotational spectrum is available in the CDMS \citep{Muller2005}.
This prediction, based in the laboratory data from \citet{McCarthy1997}, is implemented in the MADEX code \citep{Cernicharo2012} that was used
to identify the spectral features in our TMC-1 Q-band survey. \citet{McCarthy1997}
observed in the laboratory  a total of nine rotational transitions for $l$-H$_2$C$_5$
up to $J_{up}$=5 at 23 GHz. The derived parameters from \citet{McCarthy1997} (see Table \ref{rot_const})
allow to accurately predict the rotational transitions for $l$-H$_2$C$_5$ in the Q-band. The molecule has a dipole moment
of 5.9 D \citep{Maluendes1992} which makes it a very promising candidate to be observed in our
TMC-1 data. In this manner, we search for this species in our TMC-1 survey and we found a
total of eleven transitions ranging from $J_{up}$ = 7-10, whose frequencies agree very well with those predicted, discrepancies are smaller than 25 kHz. $9_{0,9}-8_{0,8}$ line is blended with a negative feature produced in the folding of the frequency switching data while $10_{0,10}-9_{0,9}$ transition is not detected due to the limited sensitivity at the predicted frequency. In any case the non-detection is consistent with the expected intensity. All the $l$-H$_2$C$_5$ lines observed
in TMC-1, shown in Fig.~\ref{spectra_h2c5} and listed in Table~\ref{table_fits}, were analyzed, together with those observed in laboratory, using an asymmetric rotor Hamiltonian with the FITWAT code \citep{Cernicharo2018} to derive the rotational and centrifugal distortion
constants. The results from this fit are shown in Table \ref{rot_const} together with those
obtained by \citet{McCarthy1997} only with laboratory data. With this new global fit an
improvement of the uncertainty in the rotational and distortion constants is obtained.
Hence, this fit is recommended to predict the frequency of the rotational transitions
of $l$-H$_2$C$_5$ with uncertainties between 10 and 200 kHz in the 50-116 GHz frequency range.

As $l$-H$_2$C$_5$ has C$_{\rm 2v}$ symmetry, it is necessary to discern between
ortho-$l$-H$_2$C$_5$ and para-$l$-H$_2$C$_5$. The ortho levels are described by $K_a$
odd while the para levels by $K_a$ even. The nuclear spin-weights are 3 and 1 for
ortho-$l$-H$_2$C$_5$ and para-$l$-H$_2$C$_5$, respectively. The $J_{K_{\rm a},K_{\rm c}}$
= 1$_{1,1}$ is the lowest ortho energy state located 13.4 K above the para ground level,
$J_{K_{\rm a},K_{\rm c}}$ = 0$_{\rm 0,0}$. Hence, from the eleven observed lines eight
of them correspond to ortho-$l$-H$_2$C$_5$ and the remaining three to para-$l$-H$_2$C$_5$.
An analysis of the line intensities through a line model fitting procedure \citet{Cernicharo2021a}
provides a rotational temperature of $\sim\,10$\,K and column densities
N(ortho-$l$-H$_2$C$_5$)=(1.3$\pm$0.3)$\times$10$^{10}$cm$^{-1}$ and N(para-$l$-H$_2$C$_5$)=(5.0$\pm$2.0)$\times$10$^{9}$cm$^{-1}$. The ortho/para ratio is
calculated to be 2.6$\pm$1.5. We have assumed a linewidth of 0.6 km s$^{-1}$ and a source of uniform brightness temperature with a diameter of 80$''$ \citep{Fosse2001}. Figure \ref{spectra_h2c5} shows, in red, the computed synthetic spectrum.

\begin{table}
\tiny
\caption{New derived rotational parameters (in MHz) for $l$-H$_2$C$_5$.}
\label{rot_const}
\centering
\begin{tabular}{{lcc}}
\hline
\hline
Constant          &  TMC-1+Lab$^a$             & Lab$^b$   \\
\hline
$A$               &  277600.0$^c$             &    277600.0       \\
$B$               &    2304.78432(27)       &      2304.7844(3) \\
$C$               &    2285.80518(27)       &      2285.8053(3) \\
$\Delta_{J}\times10^{3}$      &     0.0978(19)         &       0.104(6)    \\
$\Delta_{JK}$     &    0.04647(19)          &       0.0464(2)   \\
$rms$$^d$         &    8.7                    &       3.0         \\
$J_{min}/J_{max}$ &    2/10                   &       2/5         \\
$K_{min}/K_{max}$ &    0/1                    &       0/1         \\
N$^e$             &    20                     &       9           \\
\hline
\end{tabular}
\tablefoot{
	\tablefoottext{a}{Fit to the lines of $l$-H$_2$C$_5$ observed in TMC-1 and in laboratory \citep{McCarthy1997}.}
	\tablefoottext{b}{\citep{McCarthy1997}.}
	\tablefoottext{c}{Fixed to the value reported by \citet{McCarthy1997}.}
	\tablefoottext{d}{The standard deviation of the fit in kHz.}
	\tablefoottext{e}{Number of lines included in the fit.}
    }
\end{table}
\normalsize

\subsection{Column densities for $l$-H$_2$C$_3$, $l$-H$_2$C$_4$ and $l$-H$_2$C$_6$ in TMC-1}
\label{col_den}

\begin{table*}
\begin{center}
\caption{Observed lines of $l$-H$_2$C$_3$, $l$-H$_2$C$_4$, $l$-H$_2$C$_5$, and $l$-H$_2$C$_6$ towards TMC-1.}
\label{table_fits}
\resizebox{.60\textwidth}{!}{
\begin{tabular}{cclrrlllr}
\hline\hline
State & Transition  & \multicolumn{1}{c}{Rest Freq.} & \multicolumn{1}{c}{$E_{\rm up}$}  & \multicolumn{1}{c}{$S_{\rm ij}$} & \multicolumn{1}{c}{$\int T_{\rm A}^* dv$} & \multicolumn{1}{c}{$V_{\rm LSR}$} & \multicolumn{1}{c}{$\Delta$v} & \multicolumn{1}{c}{$T_{\rm A}^*$}\\
& $(J_{K_{\rm a},K_{\rm c}})_{\rm u}-(J_{K_{\rm a},K_{\rm c}})_{\rm l}$ & \multicolumn{1}{c}{(MHz)}      & \multicolumn{1}{c}{(K)}  &
                              &  \multicolumn{1}{c}{(mK km s$^{-1}$)}  & \multicolumn{1}{c}{(km s$^{-1}$)} & \multicolumn{1}{c}{(km s$^{-1}$)} & \multicolumn{1}{c}{(mK)}      \\
\hline
\multicolumn{9}{c}{$l$-H$_2$C$_3$}\\
\hline
\textit{ortho} & $2_{1,2}-1_{1,1}$  &   41198.335(2) &   2.0 &  1.50 &  72.0208(2) & 5.71(1) & 0.61(1) & 110.3(3) \\
\textit{para}  & $2_{0,2}-1_{0,1}$  &   41584.675(1) &   3.0 &  2.00 &  50.0199(4) & 5.75(1) & 0.58(1) & 81.11(3) \\
\textit{ortho} & $2_{1,1}-1_{1,0}$  &   41967.671(2) &   2.0 &  1.50 &  72.2521(2) & 5.75(1) & 0.56(1) & 122.5(3) \\
\hline
\multicolumn{9}{c}{$l$-H$_2$C$_4$}\\
\hline
\textit{ortho} & $4_{1,4}-3_{1,3}$  &   35577.008(2) &   3.8 &  3.75 &  137.590(1) & 5.71(1) & 0.70(1) & 185.3(2) \\
\textit{para}  & $4_{0,4}-3_{0,3}$  &   35727.379(1) &   4.3 &  4.00 &  91.6939(2) & 5.53(1) & 0.67(1) & 128.0(2) \\
\textit{ortho} & $4_{1,3}-3_{1,2}$  &   35875.775(2) &   3.9 &  3.75 &  140.130(1) & 5.69(1) & 0.69(1) & 189.9(2) \\
\textit{ortho} & $5_{1,5}-4_{1,4}$  &   44471.137(2) &   6.0 &  4.80 &  120.160(1) & 5.72(1) & 0.57(1) & 199.4(4) \\
\textit{para}  & $5_{0,5}-4_{0,4}$  &   44659.015(1) &   6.0 &  5.00 &  73.0765(2) & 5.58(1) & 0.58(1) & 119.2(4) \\
\textit{ortho} & $5_{1,4}-4_{1,3}$  &   44844.590(2) &   6.4 &  4.80 &  117.010(1) & 5.70(1) & 0.59(1) & 186.6(4) \\
\hline
\multicolumn{9}{c}{$l$-H$_2$C$_5$}\\
\hline
\textit{ortho} & $7_{1,7}-6_{1,6}$  &   32066.902(2) &   5.9 &  6.86 &  0.6759(5) & 5.83(7) & 0.69(17) & 0.9(2) \\
\textit{para}  & $7_{0,7}-6_{0,6}$  &   32133.937(2) &   6.2 &  7.00 &  1.3465(2) & 5.83(8) & 1.06(16) & 1.2(2) \\
\textit{ortho} & $7_{1,6}-6_{1,5}$  &   32199.756(2) &   6.0 &  6.86 &  1.1423(2) & 5.83(5) & 0.77(12) & 1.4(2) \\
\textit{ortho} & $8_{1,8}-7_{1,7}$  &   36647.836(3) &   7.7 &  7.88 &  0.7123(5) & 5.71(8) & 0.58(13) & 1.2(3) \\
\textit{para}  & $8_{0,8}-7_{0,7}$  &   36724.433(3) &   7.9 &  8.00 &  0.5152(11)& 5.83(8) & 0.53(23) & 0.9(3) \\
\textit{ortho} & $8_{1,7}-7_{1,6}$  &   36799.669(3) &   7.7 &  7.88 &  0.5908(7) & 5.83(5) & 0.48(19) & 1.2(2) \\
\textit{ortho} & $9_{1,9}-8_{1,8}$  &   41228.749(4) &   9.7 &  8.89 &  0.7447(6) & 5.83(8) & 0.54(16) & 1.2(4) \\
\textit{para}  & $9_{0,9}-8_{0,8}$  &   41314.902(5) &   9.9 &  9.00 &  -$^{(a)}$       & ...     & ...      & ... \\
\textit{ortho} & $9_{1,8}-8_{1,7}$  &   41399.562(4) &   9.7 &  8.89 &  0.4890(14)& 5.83(9) & 0.59(29) & 0.8(4) \\
\textit{ortho} & $10_{1,10}-9_{1,9}$ &  45809.639(6) &  11.9 &  9.90 &  0.5462(8) & 5.83(7) & 0.37(11) & 1.4(4) \\
\textit{para}  & $10_{0,10}-9_{0,9}$ &  45905.342(7) &  12.1 &  10.0 &  -$^{(b)}$       & ...     & ...      & ... \\
\textit{ortho} & $10_{1, 9}-9_{1,8}$ &  45999.431(6) &  11.9 &  9.90 &  0.6841(5) & 5.83(7) & 0.42(10) & 1.5(4) \\
\hline
\multicolumn{9}{c}{$l$-H$_2$C$_6$}\\
\hline
\textit{ortho} & $12_{1,12}-11_{1,11}$  &   32232.270(6) &    9.9 &  11.9 &  3.89347(2) & 5.82(2) & 0.80(5)  & 4.6(2) \\
\textit{para}  & $12_{0,12}-11_{0,11}$  &   32273.061(6) &   10.1 &  12.0 &  2.25233(5) & 5.86(3) & 0.73(7)  & 2.9(2) \\
\textit{ortho} & $12_{1,11}-11_{1,10}$  &   32313.117(6) &   10.0 &  11.9 &  4.17934(2) & 5.83(2) & 0.81(5)  & 4.8(3) \\
\textit{ortho} & $13_{1,13}-12_{1,12}$  &   34918.254(8) &   11.6 &  12.9 &  3.58530(2) & 5.84(2) & 0.66(4)  & 5.1(3) \\
\textit{para}  & $13_{0,13}-12_{0,12}$  &   34962.439(8) &   11.7 &  13.0 &  2.61780(6) & 5.86(4) & 0.91(9)  & 2.7(3) \\
\textit{ortho} & $13_{1,12}-12_{1,11}$  &   35005.838(8) &   11.6 &  12.9 &  3.52717(2) & 5.83(1) & 0.70(4)  & 4.7(2) \\
\textit{ortho} & $14_{1,14}-13_{1,13}$  &   37604.22(1)  &   13.4 &  13.9 &  2.95633(3) & 5.80(2) & 0.67(4)  & 4.1(2) \\
\textit{para}  & $14_{0,14}-13_{0,13}$  &   37651.80(1)  &   13.6 &  14.0 &  1.73602(10) & 5.83(4) & 0.70(9)  & 2.3(3) \\
\textit{ortho} & $14_{1,13}-13_{1,12}$  &   37698.54(1)  &   13.4 &  13.9 &  2.87938(3) & 5.80(2) & 0.73(5)  & 3.7(2) \\
\textit{ortho} & $15_{1,15}-14_{1,14}$  &   40290.19(1)  &   15.3 &  14.9 &  2.58873(5) & 5.75(2) & 0.56(5)  & 4.4(4) \\
\textit{para}  & $15_{0,15}-14_{0,14}$  &   40341.16(1)  &   15.5 &  15.0 &  1.95729(10) & 5.87(5) & 0.76(9)  & 2.4(4) \\
\textit{ortho} & $15_{1,14}-14_{1,13}$  &   40391.25(1)  &   15.4 &  14.9 &  2.47566(5) & 5.91(3) & 0.65(8)  & 3.6(2) \\
\textit{ortho} & $16_{1,16}-15_{1,15}$  &   42976.15(2)  &   17.4 &  15.9 &  1.2139(2) & 5.92(4) & 0.48(7)  & 2.4(3) \\
\textit{para}  & $16_{0,16}-15_{0,15}$  &   43030.51(2)  &   17.6 &  16.0 &  0.7463(4) & 5.80(6) & 0.50(12) & 1.4(3) \\
\textit{ortho} & $16_{1,15}-15_{1,14}$  &   43083.94(2)  &   17.4 &  15.9 &  1.61939(11) & 5.88(3) & 0.62(8)  & 2.5(3) \\
\textit{ortho} & $17_{1,17}-16_{1,16}$  &   45662.09(2)  &   19.6 &  16.9 &  1.0792(4) & 5.86(9) & 0.68(23) & 1.5(5) \\
\textit{para}  & $17_{0,17}-16_{0,16}$  &   45719.84(2)  &   19.7 &  17.0 &  1.3389(2) & 5.92(6) & 0.61(12) & 2.1(4) \\
\textit{ortho} & $17_{1,16}-16_{1,15}$  &   45776.62(2)  &   19.6 &  16.9 &  0.6063(12) & 5.85(9) & 0.45(24) & 1.3(6) \\
\textit{ortho} & $18_{1,18}-17_{1,17}$  &   48348.02(2)  &   21.9 &  17.9 &  1.3459(3) & 5.72(9) & 0.68(19) & 1.8(6) \\
\textit{para}  & $18_{0,18}-17_{0,17}$  &   48409.16(2)  &   22.1 &  18.0 &  0.36152(11) & 5.82(9) & 0.23(90) & 1.4(6) \\
\textit{ortho} & $18_{1,17}-17_{1,16}$  &   48469.29(2)  &   22.0 &  17.9 &  0.37677(10) & 6.27(5) & 0.23(90) & 1.5(6) \\
\hline
\end{tabular}
}
\end{center}
\tablefoot{Numbers in parentheses indicate the uncertainty in units of the last significant digits. For the observational parameters we adopted the uncertainty of the Gaussian fit provided by \texttt{GILDAS}. \tablefoottext{a}{One-channel spectral feature.} \tablefoottext{b}{Spectral feature below the noise level.}\\
}

\end{table*}

\begin{figure}
\centering
\includegraphics[angle=0,width=0.25\textwidth]{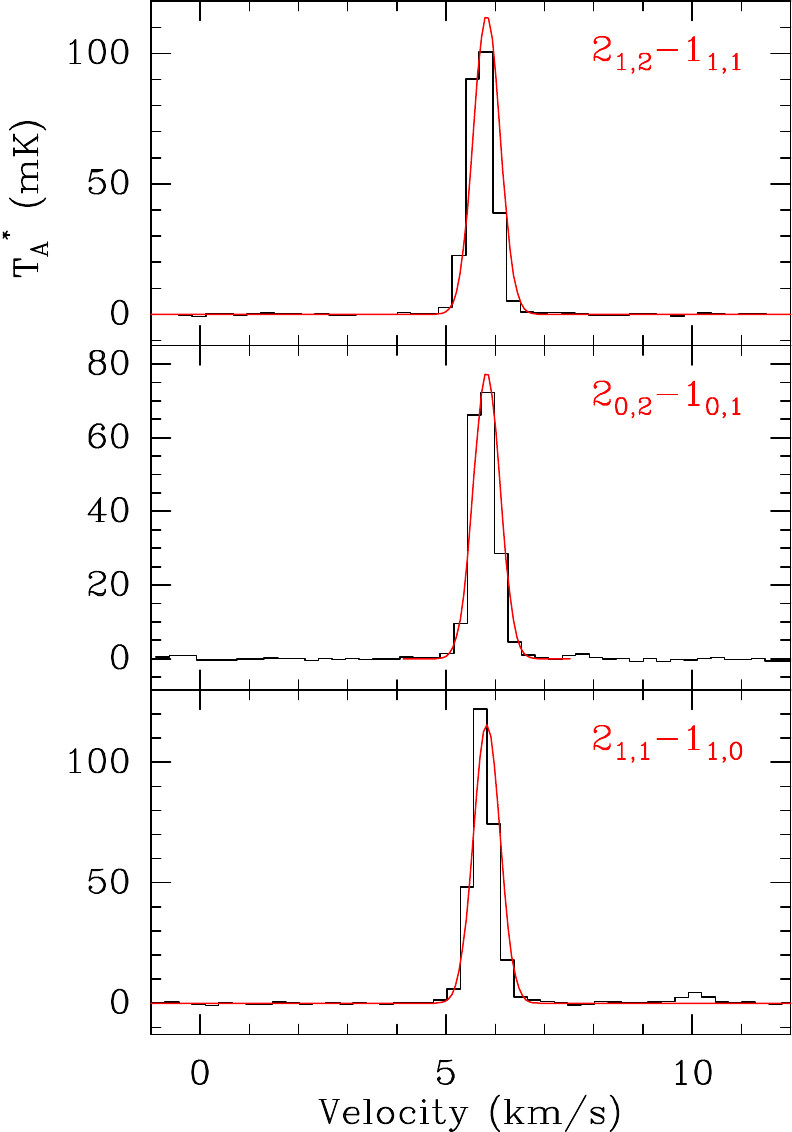}
\caption{Observed lines of $l$-H$_2$C$_3$ in TMC-1 in the 31.0-50.4 GHz range. Curves shown in red are
the computed synthetic spectra. Frequencies and line parameters are given in Table \ref{table_fits}.} \label{spectra_h2c3}
\end{figure}

The carbene $l$-H$_2$C$_3$ is the smallest cumulene species studied
in this work. The laboratory spectroscopic data used to predict the $l$-H$_2$C$_3$ spectrum
were reported by \citet{Vrtilek1990}. The electric dipole moment calculated for this
molecule is 4.1 D \citep{Defrees1986}. As for $l$-H$_2$C$_5$, $l$-H$_2$C$_3$ has
C$_{\rm 2v}$ symmetry and, thus, it is necessary to discern between ortho and
para $l$-H$_2$C$_3$. In the same manner than for $l$-H$_2$C$_5$ the ortho levels
are described by $K_a$ odd while the para levels by $K_a$ even, with 3/1 ratio
for ortho/para. Only three rotational transitions for $l$-H$_2$C$_3$ lie within the 31.0-50.4 GHz frequency range. One transition, 2$_{0,2}$-1$_{0,1}$, corresponds to
para-$l$-H$_2$C$_3$ and the other two, 2$_{1,2}$-1$_{1,1}$ and 2$_{1,1}$-1$_{1,0}$  are
ortho-$l$-H$_2$C$_3$ transitions. The lines are shown in Fig.\ref{spectra_h2c3} while the
line parameters are collected in Table \ref{table_fits}. The position of the lines is consistent
with the calculated frequencies and the systemic velocity of the source,
$V_{\rm LSR}$ = 5.83 km s$^{-1}$ \citep{Cernicharo2020}. We assumed $T_{\rm r}=10$\,K and
the column densities N(ortho-$l$-H$_2$C$_3$)=(1.5$\pm$0.5)$\times$10$^{12}$cm$^{-1}$
and N(para-$l$-H$_2$C$_3$)=(4.0$\pm$1.2)$\times$10$^{11}$cm$^{-1}$. The ortho/para
ratio is calculated to be 3.8$\pm$1.1. Collisional rates are available for the
system l-H$_2$C$_3$/He \citep{Khalifa2019}. Adopting a volume density for TMC-1 of
4$\times$10$^4$\,cm$^-{3}$ \citep{Cernicharo1987,Fosse2001} we derive an excitation temperature for the two ortho
transitions of $\sim$9\,K, and of $\sim$8.5\,K for the para transition. Hence, the adopted
rotational temperature seems well adapted to the excitation conditions of this molecule.

\begin{figure}
\centering
\includegraphics[angle=0,width=\columnwidth]{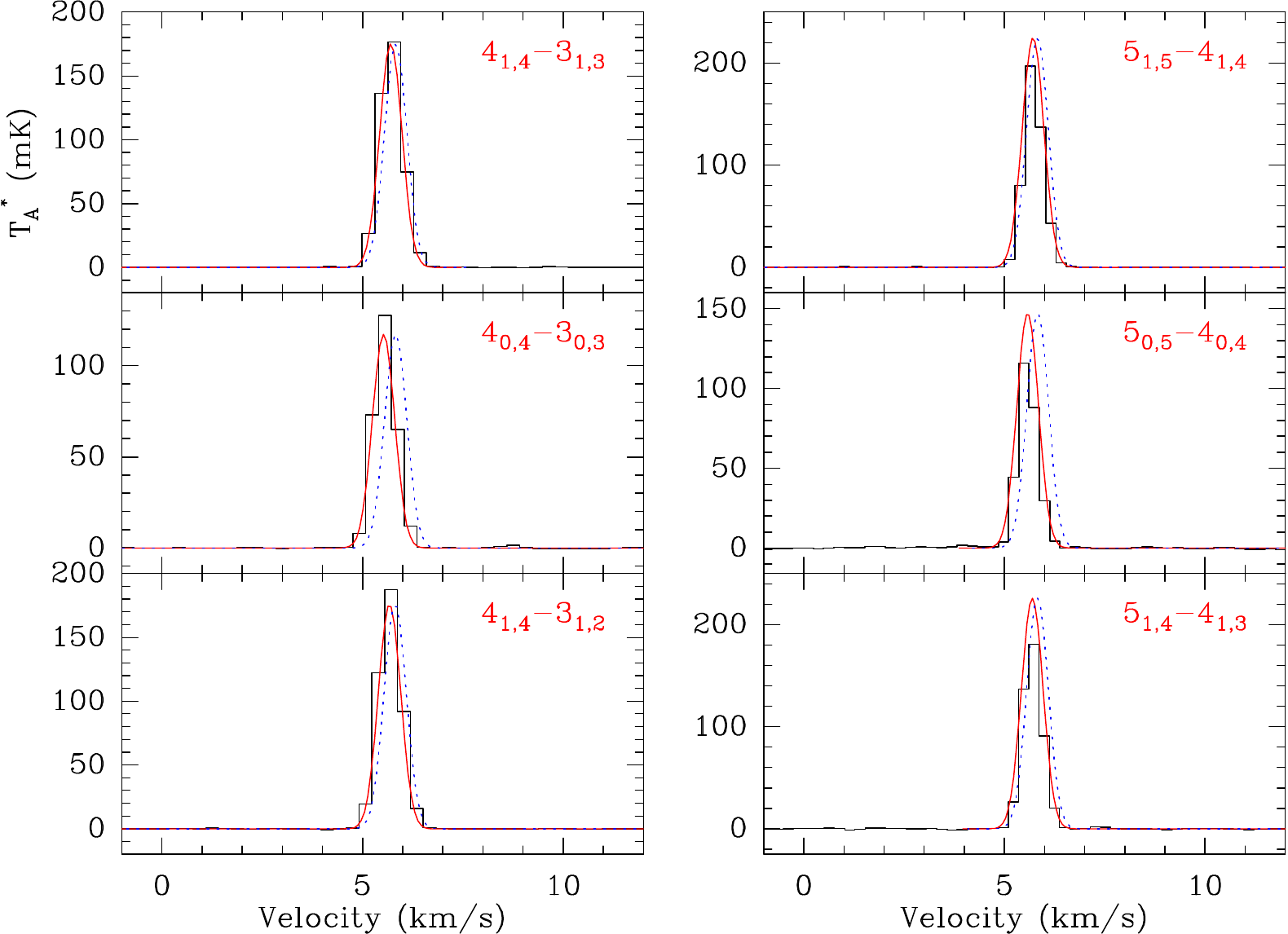}
\caption{Observed lines of $l$-H$_2$C$_4$ in TMC-1 in the 31.0-50.4 GHz range. Curves shown in red and green are the are the computed synthetic spectra using the observed frequencies and the CDMS predictions, respectively. Frequencies and line parameters are given in Table \ref{table_fits}.} \label{spectra_h2c4}
\end{figure}

The next member of the $l$-H$_2$C$_n$ family is $l$-H$_2$C$_4$. The rotational spectrum
for this cumulene was observed by \citet{Killian1990} and \citet{Travers1996}.
From ab initio calculations the dipole moment of $l$-H$_2$C$_4$ has been estimated
to be 4.1\,D \citep{Oswald1995}, similar to the smaller cumulene carbene $l$-H$_2$C$_3$. The lines for ortho- and para-$l$-H$_2$C$_4$ are clearly
detected in our TMC-1 survey as can be seen in Fig. \ref{spectra_h2c4}. However, it should be noted that there are small discrepancies (of about 30\,kHz for the para species) between the observed and predicted frequencies from CDMS. We observed a total of six rotational transitions with $J_{up}$ = 4 and 5 and $K_{\rm a}$ = 0 and 1. Four of them pertain to ortho-$l$-H$_2$C$_4$ and two to para-$l$-H$_2$C$_4$. All the line parameters are given in Table \ref{table_fits}. From the observed integrated line intensities we obtained a $T_{\rm r}=10$\,K and column densities for the ortho and para species of N(ortho-$l$-H$_2$C$_4$)=(2.5$\pm$0.8)$\times$10$^{12}$cm$^{-1}$ and N(para-$l$-H$_2$C$_4$)=(8.0$\pm$2.3)$\times$10$^{11}$cm$^{-1}$. The ortho/para ratio for $l$-H$_2$C$_4$ is 3.1$\pm$0.9.

\begin{figure}
\centering
\includegraphics[angle=0,width=\columnwidth]{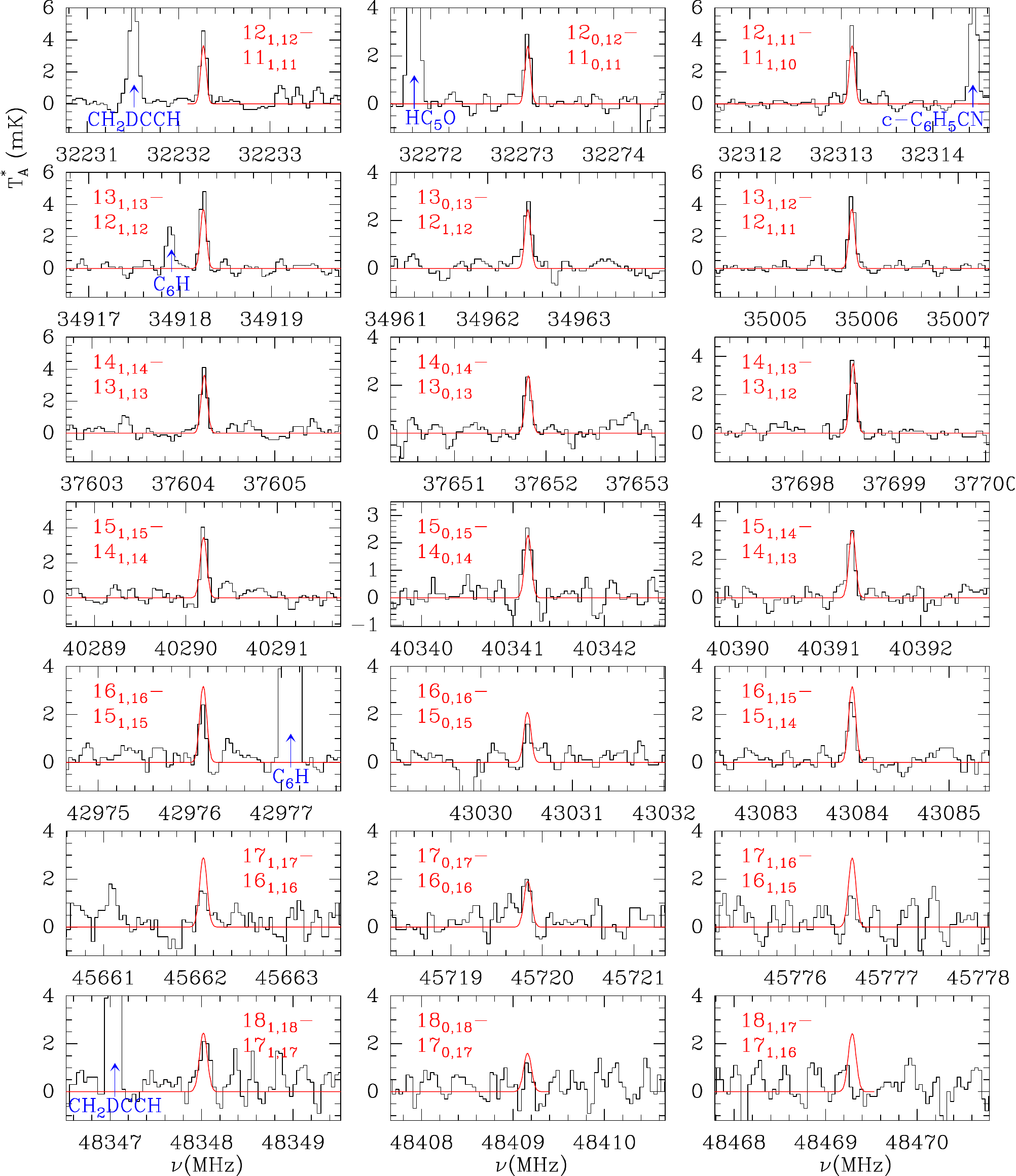}
\caption{Observed lines of $l$-H$_2$C$_6$ in TMC-1 in the 31.0-50.4 GHz range. Curves shown in red are
the computed synthetic spectra. Frequencies and line parameters are given
in Table \ref{table_fits}.} \label{spectra_h2c6}
\end{figure}

$l$-H$_2$C$_6$ is, to date, the larger cumulene carbene observed in space.
Its dipole moment is calculated \citep{Maluendes1992} to be larger than those
for the smaller cumulenes, with a value of 6.2 D. The rotational spectrum has
been investigated by \citet{McCarthy1997} and the derived spectroscopic parameters
are used to predict its transition frequencies. Due to its larger molecular
size, many rotational transition for $l$-H$_2$C$_6$ can be observed in the
31.0-50.4 GHz frequency range. As for all the other cumulene carbenes, and due to its symmetry,
it is necessary to discern between ortho- and para-$l$-H$_2$C$_6$. Hence, we observed a
total of twenty-one transitions for $l$-H$_2$C$_6$, fourteen for ortho- and seven
for para-$l$-H$_2$C$_6$ (see Fig. \ref{spectra_h2c6}). Line parameters are collected in
Table \ref{table_fits}. We derived from the observed integrated line intensities
a $T_{\rm r}=10$\,K and the column densities
N(ortho-$l$-H$_2$C$_6$)=(6.0$\pm$1.8)$\times$10$^{10}$cm$^{-1}$ and
N(para-$l$-H$_2$C$_3$)=(2.0$\pm$0.6)$\times$10$^{10}$cm$^{-1}$. The ortho/para ratio
is calculated to be 3.0$\pm$0.9.

From our observations we obtain the following relative abundances for $l$-H$_2$C$_3$/$l$-H$_2$C$_4$/$l$-H$_2$C$_5$/$l$-H$_2$C$_6$ in TMC-1,
344/561/1/4.4. $l$-H$_2$C$_4$ is the most abundant cumulene carbene in TMC-1, followed by $l$-H$_2$C$_3$. The larger species $l$-H$_2$C$_5$ and $l$-H$_2$C$_6$
are much less abundant compared to the shorter cumulene chain, with $l$-H$_2$C$_6$ being 4.4 times more abundant than the odd member $l$-H$_2$C$_5$. It is worth noting that, within the uncertainties, the four cumelenes studied in this work have an ortho/para abundance ratio of 3. Hence, no significant ortho to para conversion can be noticed for these molecules.

\subsection{Chemistry of cumulene carbenes}
\label{chem}

In order to understand how cumulene carbenes $l$-H$_2$C$_n$ can be formed in TMC-1, we carried out gas-phase chemical modeling calculations. We adopted typical conditions of cold dark clouds, i.e., a volume density of H nuclei of 2\,$\times$10$^4$ cm$^{-3}$, a gas kinetic temperature of 10 K, a visual extinction of 30 mag, a cosmic-ray ionization rate of H$_2$ of 1.3\,$\times$\,10$^{-17}$ s$^{-1}$, and the so-called set of low-metal elemental abundances (e.g., \citealt{Agundez2013}). We used the chemical network {\small RATE12} from the {\small UMIST} database \citep{McElroy2013}, updated with results from \cite{Lin2013}, and expanded with the subset of gas-phase chemical reactions revised by \cite{Loison2017} in their study of the chemistry of C$_3$H and C$_3$H$_2$ isomers.

\begin{figure}
\centering
\includegraphics[angle=0,width=\columnwidth]{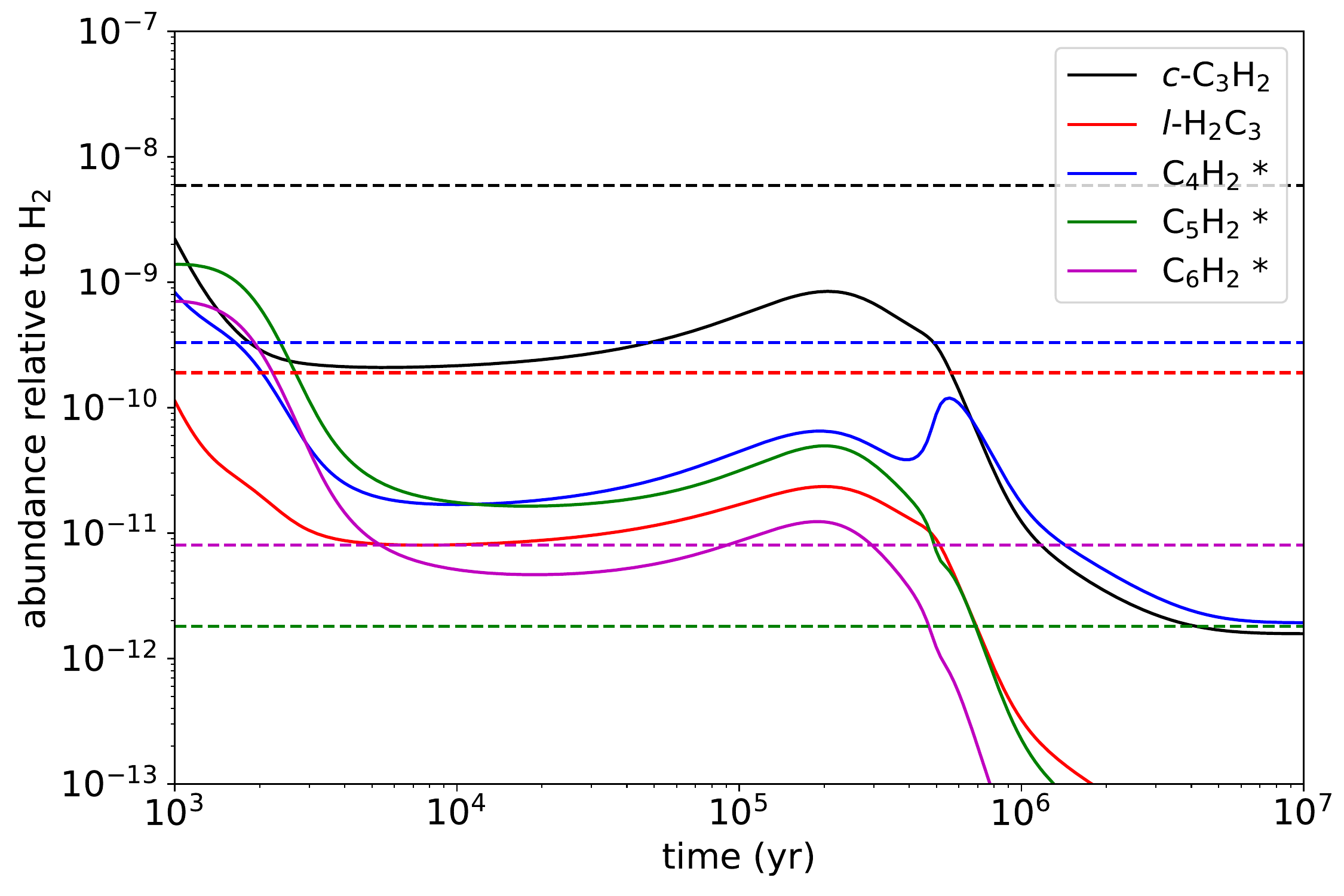}
\caption{Calculated fractional abundances of various hydrocarbons as a function of time. The symbol $*$ in the legend means that the chemical model does not distinguish between different isomers. The abundances observed
in TMC-1 for $c$-C$_3$H$_2$ \citep{Cernicharo2021a} and $l$-H$_2$C$_n$ with $n$\,=\,3-6 (this work) are indicated by dashed horizontal lines.} \label{fig:abun}
\end{figure}

Among the family of carbenes $l$-H$_2$C$_n$, the one for which the chemistry is better constrained is by far the smallest member $l$-H$_2$C$_3$, which has
been discussed in detail by \cite{Loison2017}. This species is mainly formed upon dissociative recombination with electrons of the linear and cyclic isomers of C$_3$H$_3^+$, which in turn are formed through the radiative association of C$_3$H$^+$ and H$_2$. If we focus in the so-called early time, a few 10$^5$ yr, where gas-phase chemical models of cold dark clouds reproduce better TMC-1 observations (e.g., \citealt{Agundez2013}), the calculated abundance is about one order of magnitude below the observed value (see Fig.~\ref{fig:abun}), but the calculated cyclic-to-linear abundance ratio agrees very well with the observed value of 31 (\citealt{Cernicharo2021a}; this work). For members of the series $l$-H$_2$C$_n$ with $n$\,$>$\,3, information on the chemistry of the different possible isomers is poorly known and thus chemical networks, such as {\small UMIST RATE12}, do not distinguish between them. The calculated abundances of C$_4$H$_2$ and C$_6$H$_2$ agree within a factor of 2-3 with the observed abundances of $l$-H$_2$C$_4$ and $l$-H$_2$C$_6$, respectively (see Fig.~\ref{fig:abun}), although one must keep in mind that observations only refer to the cumulene while
the model includes also other isomers, in particular the more stable non-polar species HC$_n$H. If the isomer HC$_n$H is significantly more abundant than H$_2$C$_n$ for n = 4,6 in TMC-1, then the abundances calculated by the chemical model for C$_4$H$_2$ and C$_6$H$_2$ would be too low. In the case of C$_5$H$_2$, the calculated peak abundance is about 30 times lower than the observed abundance of $l$-H$_2$C$_5$ (see Fig.~\ref{fig:abun}), although this may not be a problem if the more stable isomer HC$_5$H is substantially more abundant than the carbene $l$-H$_2$C$_5$.

For $l$-H$_2$C$_n$ with $n$\,$>$\,3, the chemical route analogous to that forming $l$-H$_2$C$_3$ has variable degrees of efficiency. For example, formation of C$_4$H$_3^+$ is relatively efficient thanks to the radiative association between C$_4$H$_2^+$ and H \citep{McEwan1999}. For $n$\,=\,5, the route does not work because C$_5$H$^+$ does not react with H$_2$ \citep{McElvany1987,Bohme1990}, while formation of C$_6$H$_3^+$ is uncertain due to the unknown reactivity of C$_6$H$_2^+$ with H$_2$ \citep{Anicich2003}.
Further studies of reactions involving hydrocarbon ions, in particular regarding isomer differentiation, are highly desirable. According to the chemical model, reactions of C$_n$H$^-$ anions with H atoms are a major route to C$_n$H$_2$ molecules, such as C$_4$H$_2$, C$_5$H$_2$, and C$_6$H$_2$. These reactions have been studied in the laboratory for anions C$_n$H$^-$ with $n$\,=\,2, 4, 6, and 7 and have been found to be rapid and to yield C$_n$H$_2$ as main product \citep{Barckholtz2001}. The reaction with $n$\,=\,5, although not studied, is likely to behave similarly. It is however unknown whether the carbene isomer H$_2$C$_n$ or the more stable HC$_n$H is preferentially formed. It would be very helpful to investigate this particular point.

Other formation routes, apart from C$_n$H$_m^+$ + e$^-$ and C$_n$H$^-$ + H, can be provided by neutral-neutral reactions. For example, the chemical model points to the reaction between atomic C and the propargyl radical (CH$_2$CCH), recently detected in TMC-1 \citep{Agundez2021}, as a source of C$_4$H$_2$ isomers. This reaction is assumed to be fast by \cite{Loison2017}, but calculations of the rate coefficient and product distribution at low temperatures are needed. Similarly, the reaction C + C$_4$H$_3$ is
assumed to proceed fast by \cite{Smith2004} and provides an important route to C$_5$H$_2$ isomers, but more detailed studies on this reaction are necessary. Finally, we note that the cyclic C$_5$H$_2$ isomer $c$-C$_3$HCCH recently detected by \citet{Cernicharo2021a} is most likely formed through the reaction between CCH and $c$-C$_3$H$_2$.

\begin{acknowledgements}
This research has been funded by ERC through grant ERC-2013-Syg-610256-NANOCOSMOS. Authors also thank Ministerio de Ciencia e Innovaci\'on for funding support through projects AYA2016-75066-C2-1-P, PID2019-106235GB-I00 and PID2019-107115GB-C21 / AEI / 10.13039/501100011033. MA thanks Ministerio de Ciencia e Innovaci\'on for grant RyC-2014-16277.

\end{acknowledgements}

\end{document}